\documentstyle[12pt]{article}
\begin{document}
\begin{center}
{\Large \bf  Quantum Gravity and Black Hole Entropy}
\vskip 0.3 true in
{\large J. W. Moffat}
\date{}
\vskip 0.3 true in
{\it Department of Physics, University of Toronto,
Toronto, Ontario M5S 1A7, Canada}
\vskip 0.3 true in
Invited Talk presented at the XI International Conference on Problems in Quantum Field
Theory, July 13-17, 1998.
\vskip 0.3 true in
\end{center}

\begin{abstract}%
The basic features of a quantum field theory which is Poincar\'e invariant, gauge invariant, finite
and unitary to all orders of perturbation theory are reviewed. Quantum gravity is finite and
unitary to all orders of perturbation theory. The Bekenstein-Hawking entropy formula
for a black hole is investigated in a conical Rindler space approximation to a black hole event
horizon. A renormalization of the gravitational coupling constant is performed leading 
to a finite Bekenstein-Hawking entropy at the horizon.
\end{abstract}

\section{\bf Finite Quantum Field Theory}

A finite quantum field theory (FQFT) based on a nonlocal interaction Lagrangian has been
developed which is perturbatively finite, unitary and gauge invariant [1-9]. The finiteness draws
from the fact that factors of $\exp[{\cal K}(p^2)/2\Lambda_F^2]$ are attached to propagators
which suppress any ultraviolet divergences in Euclidean momentum space.

An important development in FQFT was the discovery that gauge invariance and
unitarity can be restored by adding series of higher interactions. The
resulting theory possesses a nonlinear, field representation dependent gauge invariance which
agrees with the original
local symmetry on shell but is larger off shell. Quantization is performed in the functional
formalism using an analytic and convergent measure factor which retains invariance
under the new symmetry. An explicit calculation was made of the measure factor in
QED\protect\cite{Moffat2}, and it was obtained to lowest order in Yang-Mills
theory\protect\cite{Kleppe2}. Kleppe and Woodard\protect\cite{Woodard2} obtained an ansatz
based on the derived dimensionally regulated result when
$\Lambda_F\rightarrow\infty$, which was conjectured to lead to a general functional measure
factor in FQFT gauge theories. 

A convenient formalism which makes the FQFT construction transparent is based on shadow
fields\protect\cite{Kleppe2,Woodard2}. Let us denote by $f_i$ a generic local field and write the
standard local action as
\begin{equation}
W[f]=W_F[f]+W_I[f],
\end{equation}
where $W_F$ and $W_I$ denote the free part and the interaction part
of the action, respectively, and
\begin{equation}
W_F=\frac{1}{2}\int d^4xf_i(x){\cal K}_{ij}f_j(x).
\end{equation}
In a gauge theory $W$ would be the Becchi, Rouet, Stora (BRS) gauge-fixed action including
ghost fields in the invariant action required to fix the gauge.  The kinetic operator ${\cal K}$ is
fixed by defining a Lorentz-invariant distribution operator:
\begin{equation}
{\cal E}\equiv \exp\biggl(\frac{{\cal K}}{2\Lambda_F^2}\biggr)
\end{equation}
and the shadow operator:
\begin{equation}
{\cal O}^{-1}=\frac{{\cal K}}{{\cal E}^2-1},
\end{equation}
where $\Lambda_F$ is an energy scale parameter.

Every field $f_i$ has an auxiliary counterpart field $h_i$, and they are used to form a new action:
\begin{equation}
W[f,h]\equiv W_F[F]-A[h]+W_I[f+h],
\end{equation}
where
\[
F={\cal E}^{-1}f,\quad
A[h]=\frac{1}{2}\int d^4xh_i{\cal O}^{-1}_{ij}h_j.
\]
By iterating the equation
\begin{equation}
h_i={\cal O}_{ij}\frac{\delta W_I[f+h]}{\delta h_j}
\end{equation}
the shadow fields can be determined as functionals, and the regulated action is derived from
\begin{equation}
\hat W[f]=W[f,h(f)].
\end{equation}
We recover the original local action when we take the limit $\Lambda_F\rightarrow\infty$ and
$\hat f\rightarrow f, h(f)\rightarrow 0$.

Quantization is performed using the definition
\begin{equation}
\langle 0\vert T^*(O[f])\vert 0\rangle_{\cal E}=\int[Df]\mu[f]({\rm gauge\, fixing})
O[F]\exp(i\hat W[f]).
\end{equation}
On the left-hand side we have the regulated vacuum expectation value of the
$T^*$-ordered product of an arbitrary operator $O[f]$ formed from the local fields $f_i$. The
subscript ${\cal E}$ signifies that a
regulating Lorentz distribution has been used. Moreover, $\mu[f]$ is a measure factor and there
is a gauge fixing factor, both of which are needed to maintain perturbative unitarity in
gauge theories.

The new Feynman rules for FQFT are obtained as follows: The vertices remain unchanged but
every leg of a diagram is connected either to a regularized propagator,
\begin{equation}
\label{regpropagator}
\frac{i{\cal E}^2}{{\cal K}+i\epsilon}
=-i\int^{\infty}_1\frac{d\tau}{\Lambda_F^2}\exp\biggl(\tau
\frac{{\cal K}}{\Lambda^2_F}\biggr),
\end{equation}
or to a shadow propagator,
\begin{equation}
-i{\cal O}=\frac{i(1-{\cal E}^2)}{{\cal K}}=-i\int^1_0\frac{d\tau}{\Lambda_F^2}
\exp\biggl(\tau\frac{{\cal K}}{\Lambda_F^2}\biggr).
\end{equation}
The formalism is set up in Minkowski spacetime and loop integrals are formally
defined in Euclidean space by performing a Wick rotation. This fascilitates the analytic
continuation; the whole formalism could from the outset be developed in Euclidean space.

In FQFT renormalization is carried out as in any other field theory. The bare parameters are
calculated from the renormalized ones and $\Lambda_F$, such that the limit
$\Lambda_F\rightarrow\infty$ is finite for all noncoincident Green's functions, and the bare
parameters are those of the local theory. The regularizing interactions {\it are determined by the
local operators.}

The regulating Lorentz distribution function ${\cal E}$ must be chosen to perform an explicit
calculation in perturbation theory. We do not know the unique choice of ${\cal E}$.
It maybe that there exists an equivalence mapping between all the possible
distribution functions ${\cal E}$. However, once a choice for the function is made, then the
theory and the perturbative calculations are uniquely fixed. A standard
choice in early FQFT papers is\protect\cite{Moffat,Moffat2}:
\begin{equation}
\label{reg}
{\cal E}_m=\exp\biggl(\frac{\Box-m^2}{2\Lambda_F^2}\biggr).
\end{equation}

An explicit construction for QED was given using the Cutkosky rules as applied to
FQFT whose propagators have poles only where ${\cal K}=0$ and whose vertices are entire
functions of ${\cal K}$. The regulated action $\hat W[f]$ satisfies these requirements which
guarantees unitarity on the physical space of states. The local action is gauge fixed and then a
regularization is performed on the BRS theory. 

The infinitesimal transformation
\begin{equation}
\delta f_i=T_i(f)
\end{equation}
generates a symmetry of $W$, and the infinitesimal transformation 
\begin{equation}
\hat\delta f_i={\cal E}^2_{ij}T_j(f+h[f])
\end{equation}
generates a symmetry of the regulated action ${\hat W}$. It follows that FQFT regularization
preserves all
continuous symmetries including supersymmetry. The quantum theory will preserve symmetries
provided a suitable measure factor can be found such that
\begin{equation}
\hat\delta([Df]\mu[f])=0.
\end{equation}
Moreover, the interaction vertices of the measure factor must be entire functions of the operator
${\cal K}$ and they must not destroy the FQFT finiteness.

In FQFT tree order, Green's functions remain local except for external lines which are unity on
shell. It follows immediately that since on-shell tree amplitues are unchanged by the
regularization, $\hat W$ preserves all symmetries of $W$ on shell. Also all loops
contain at least one regularizing propagator and therefore are ultraviolet finite. Shadow fields are
eliminated at the classical level, for functionally integrating over them would produce
divergences from shadow loops. Since shadow field propagators do not contain any poles there is
no need to quantize the shadow fields. Feynman rules for ${\hat W}[f,h]$ are as simple as those
for local field theory.

Kleppe and Woodard\cite{Woodard2} have calculated FQFT for $\phi^4$ scalar field theory in
four dimensions and $\phi^3$ in six dimensions to two-loop order, and explicitly shown that
problems such as overlapping divergences can be dealt with correctly. There
are no problems with power counting and the scalar field theory case is not more difficult to
implement than dimensional regularization. Indeed, one obtains the FQFT regulated result in the
limit $\Lambda_F\rightarrow\infty$ by replacing the gamma function of the dimensional
regularization result by an incomplete gamma function, evaluated as a simple combination of
Feynman parameters.

We recall that regularization schemes, such as Pauli-Villars or a conventional cut-off
method, only define a finite quantum field theory for a finite value of the
cut-off at the price of violating gauge invariance.
Dimensional regularization or $\zeta$-function regularization maintain gauge 
invariance but are finite only in a complex fractional or complex dimensional space. In contrast,
FQFT is finite in a real space with $D\geq 4$ for a fixed, finite value of $\Lambda_F$,
while preserving gauge invariance under the extended gauge transformations.

Although FQFT is perturbatively unitary, at fixed loop order the scattering amplitudes appear to
violate
bounds imposed by partial wave unitarity, i.e. the projected partial waves $A_{\ell }(s)$, where
$s$ is the Mandelstam center-of-mass energy squared, grows beyond the partial wave unitarity
limit as $s\rightarrow\infty$. Because our tree graphs are the same as the local theory this occurs
only for the loop graphs. The same problem occurs in string theory for both the tree graphs and
the  loop graphs at high energies\protect\cite{Soldate}, and it has been studied by Muzinich and
Soldate\protect\cite{Muzinich}. These studies suggested that the problem
can be resolved by a resummation of the perturbation series.

Efimov\protect\cite{Efimov2} has provided a solution to the fixed perturbative loop order
unitarity problem. He considered a system of two scalar particles with mass $m$ and the
following inequalities for the upper bound on the elastic scattering amplitude $M(s,t)$:
\begin{equation}
\vert M(s,t)\vert < C(t_0)s,\quad (\vert t\vert \geq \vert t_0\vert > 0)
\end{equation}
and for the total cross section 
\begin{equation}
\sigma_{\rm tot} \leq C\vert \frac{d}{dt}\ln {\rm Im}M(s,t)\vert_{t=0}\quad 
(s\rightarrow\infty).
\end{equation}
These bounds were obtained by using the unitarity of the S-matrix on the mass shell and the
natural assumption that the imaginary part of the elastic scattering, ${\rm Im}M(s,t)$, is a
differentiable and convex down function in the neighborhood of $t=0$. The analyticity of the
elastic scattering amplitude in the Martin-Lehmann ellipse\protect\cite{Eden} and the locality of
the theory were not used in the derivation of the bounds. 

\section{\bf Finite Quantum Gravity Theory}

We shall now formulate general relativity (GR) as a FQFT. This problem has been considered
previously\protect\cite{Moffat,Moffat2,Moffat4}, and as in ref.(\cite{Moffat4}), we will
regularize the GR equations using the shadow field formalism. 

We expand the local interpolating field ${\bf g}^{\mu\nu}=\sqrt{-g}g^{\mu\nu}\quad 
(g={\rm Det}(g_{\mu\nu}))$ about any spacetime background field, e.g., we expand about
Minkowski spacetime
\begin{equation}
{\bf g}^{\mu\nu}=\eta^{\mu\nu}+\kappa\gamma^{\mu\nu}+O(\kappa^2),
\end{equation}
where $\kappa=(32\pi G)^{1/2}$. We separate the free and interacting parts of the Lagrangian
\begin{equation}
{\cal L}_{\rm grav}(g)={\cal L}^F_{\rm grav}(g)+{\cal L}^I_{\rm grav}(g).
\end{equation}
The finite regularized gravitational Lagrangian in FQFT is given by
\begin{equation}
{\hat{\cal L}}(g,s)={\cal L}^F_{\rm grav}({\hat g})
-{\cal A}_{\rm grav}(s)+{\cal L}^I_{\rm grav}(g+s),
\end{equation}
where
\begin{equation}
{\hat g}={\cal E}^{-1}g,\quad {\cal A}_{\rm grav}(s)=F(s_i{\cal O}_{ij}^{-1}s_j),
\end{equation}
and $s$ denotes the graviton shadow field. 

The quantum gravity perturbation theory is locally $SO(3,1)$ invariant (generalized, nonlinear
field representation dependent transformations), unitary and finite to all orders in a way similar 
to non-Abelian gauge theories formulated using FQFT.  At the tree graph level
all unphysical polarization states are decoupled and nonlocal effects will only occur in graviton
and graviton-matter loop graphs. Because the gravitational tree graphs are
purely local there is a well-defined classical GR limit.  The finite quantum gravity theory is
well-defined in four real spacetime dimensions (or any real spacetime of dimensions $D> 4$). 

The graviton regularized propagator in a fixed de Donder gauge\cite{Moffat4} is given by
\begin{equation}
D^{\rm grav}_{\mu\nu\lambda\rho}
=(\eta_{\mu\lambda}\eta_{\nu\rho}+\eta_{\mu\rho}\eta_{\nu\lambda}
-\eta_{\mu\nu}\eta_{\lambda\rho})\frac{-i}{(2\pi)^4}
\int d^4k\frac{{\cal E}^2(k^2)}{k^2-i\epsilon}\exp[ik\cdot(x-y)],
\end{equation}
while the shadow propagator is
\begin{equation}
D^{\rm shad}_{\mu\nu\lambda\rho}
=(\eta_{\mu\lambda}\eta_{\nu\rho}+\eta_{\mu\rho}\eta_{\nu\lambda}
-\eta_{\mu\nu}\eta_{\lambda\rho})\frac{-i}{(2\pi)^4}
\int d^4k\frac{[1-{\cal E}^2(k^2)]}{k^2-i\epsilon}\exp[ik\cdot(x-y)].
\end{equation}
In momentum space we have
\[
\frac{-i{\cal E}^2(k^2)}{k^2-i\epsilon}=-i\int^{\infty}_1\frac{d\tau}{\Lambda^2_F}
\exp\biggl(-\tau\frac{k^2}{\Lambda^2_G}\biggr),
\]
and
\[
\frac{i({\cal E}^2(k^2)-1)}{k^2-i\epsilon}=-i\int_0^1\frac{d\tau}{\Lambda^2_F}
\exp\biggl(-\tau\frac{k^2}{\Lambda^2_G}\biggr),
\]
where $\Lambda_G$ is the gravitational scale parameter.

The Einstein-Hilbert action for pure gravity is
\begin{equation}
W_{\rm grav}=\int d^4x\sqrt{-g}\frac{2}{\kappa^2}R,
\end{equation}
and for local spinless, scalar matter fields $\phi$ the action is
\begin{equation}
W_{\rm matter}=\int d^4x\sqrt{-g}(\frac{1}{2}g^{\mu\nu}\partial_\mu\phi
\partial_\nu\phi-\frac{1}{2}m^2\phi^2).
\end{equation}

We quantize by means of the path integral operation
\begin{equation}
\langle 0\vert T^*(O[g,\phi])\vert 0\rangle_{\cal E}=\int[Dg][D\phi]\mu[g,\phi]({\rm gauge\,
fixing})
O[{\hat g},{\hat\phi}]\exp(i\hat W[g,\phi]),
\end{equation}
where $\hat\phi={\cal E}^{-1}\phi$. The quantization is carried out in the functional formalism 
by finding a measure factor
$\mu[\kappa\gamma,\phi]$ to make $[D\gamma]$ invariant under the classical symmetry.
To ensure a correct gauge fixing scheme, we write $W[g,\phi]$ 
in the BRS invariant form with ghost fields; the ghost structure arises from exponentiating
the Faddeev-Popov determinant. The algebra of gauge symmetries is not
expected to close off-shell, so one needs to introduce higher ghost terms (beyond the normal
ones) into both the action and the BRS transformation. The BRS action will be
regularized directly to ensure that all the corrections to the measure factor are included.

\section{Quantum Nonlocal Behavior in FQFT}

In FQFT, it can be
argued that the extended objects that replace point particles (the latter are obtained in the limit
$\Lambda_F\rightarrow\infty$) cannot be probed because of a Heisenberg uncertainty type of
argument.  The FQFT nonlocality {\it only occurs at the quantum loop
level}, so there is no noncausal classical behavior. In FQFT the strength of a signal propagated
over an invariant interval $l^2$ outside the
light cone would be suppressed by a factor $\exp(-l^2\Lambda_F^2)$. 

Nonlocal field theories can possess non-perturbative instabilities. These
instabilities arise because of extra canonical degrees of freedom associated with higher time
derivatives. If a Lagrangian contains up to $N$ time derivatives, then the associated Hamiltonian
is linear in $N-1$ of the corresponding canonical variables and extra canonical degrees of
freedom will be generated by the higher time derivatives. The nonlocal theory can be viewed as
the limit $N\rightarrow\infty$ of an Nth derivative Lagrangian. Unless the dependence on the
extra solutions is arbitrarily choppy in the limit, then the higher derivative limit will produce
instabilities\protect\cite{Eliezer}. The condition for the smoothness of the extra solutions is that
no invertible field
redefinition exists which maps the nonlocal field equations into the local ones. String theory does
satisfy this smoothness condition as can be seen by inspection of the S-matrix tree graphs. In
FQFT the tree amplitudes agree with those of the local theory, so the smoothness condition is not
obeyed.

It was proved by Kleppe and Woodard\protect\cite{Kleppe2} that the solutions of the nonlocal
field equations in FQFT are in one-to-one correspondence with those of the original local theory.
The relation for a generic field $v_i$ is
\begin{equation}
v_i^{\rm nonlocal}={\cal E}^2_{ij}v^{\rm local}_j.
\end{equation}
Also the actions satisfy
\begin{equation}
W[v]={\hat W}[{\cal E}^2v].
\end{equation}
Thus, there are no extra classical solutions. The solutions of the regularized nonlocal
Euler-Lagrange equations are in one-to-one correspondence with those of the local action. It
follows {\it that the regularized nonlocal FQFT is free of higher derivative solutions, so FQFT
can
be a stable theory.}

Since only the quantum loop graphs in the nonlocal FQFT differ from the local field theory,
then FQFT can be viewed as a non-canonical quantization of fields which obey the local
equations of motion. Provided the functional quantization in FQFT is successful, then the theory
does maintain perturbative unitarity. 

\section{\bf Graviton Self-Energy}

The lowest order contributions to the graviton self-energy in FQFT will include the standard
graviton loops, the shadow field graviton loops, the ghost field loop contributions with their
shadow field counterparts, and the measure loop contributions. The calculated measures for 
regularized QED, first order Yang-Mills theory and all orders in $\phi^4$ and $\phi^6$ theories
lead to self-energy contributions that are controlled by an incomplete $\Gamma$ function.
For the regularized perturbative gravity theory the first order loop amplitude is
\begin{equation}
A_i=\Gamma\biggl(2-D/2,\frac{p^2}{\Lambda_G^2}\biggr)(p^2)^{D-2}F_i(D/2),\quad
(i=1,...,5)
\end{equation}
where $\Gamma(n,z)$ is the incomplete gamma function
\[
\Gamma(n,z)=\int_z^{\infty}dt t^{n-1}\exp(-t).
\]
The dimensional regularization result is obtained by the replacement
\[
\Gamma\biggl(2-D/2,\frac{p^2}{\Lambda_G^2}\biggr)\rightarrow \Gamma(2-D/2),
\]
yielding the result\protect\cite{Capper,Veltman}
\begin{equation}
A_i\sim \Gamma(2-D/2)(p^2)^{D-2}F_i(D/2)\sim \frac{1}{\epsilon}(p^2)^{D-2}
F_i(D/2),
\end{equation}
where $\epsilon=2-D/2$ and $\Gamma(n)$ is the gamma function. Whereas the dimensional
regularization result is singular in the limit $\epsilon\rightarrow 0$, the FQFT result is 
{\it finite in this limit} for a
fixed value of the parameter $\Lambda_G$, resulting in a finite graviton self-energy
amplitude $A_i$. The parameter $\Lambda_G$ could be chosen to be the Planck energy
scale, $\Lambda_G\sim 10^{19}$ GeV (in ref.(\protect\cite{Moffat4}), the universal scale
$\Lambda_F=\Lambda_G
=1/\sqrt{G_F}\sim 300$ Gev, where $G_F$ is Fermi's weak coupling constant, was chosen to
solve the Higgs gauge hierarchy problem in the standard model and stabilize the proton).

For $D=4$ spacetime we have
\[
\Gamma\biggl(0,\frac{p^2}{\Lambda_G^2}\biggr)=E_i\biggl(\frac{p^2}{\Lambda_G^2}\biggr),
\]
where $E_i(z)$ is the error function. We have in the Euclidean momentum limit
$p^2\rightarrow\infty$:
\begin{equation}
E_i\biggl(\frac{p^2}{\Lambda_G^2}\biggr)\sim \exp\biggl(-\frac{p^2}{\Lambda_G^2}\biggr)
\biggl(\frac{\Lambda_G^2}{p^2}-\frac{\Lambda_G^4}{p^4}+...\biggr).
\end{equation}
Thus, in the infinite Euclidean momentum limit the quantum graviton self-energy contribution
damps out and quantum graviton corrections become negligible. It is often argued in the
literature on quantum gravity that the gravitational quantum corrections scale as $\alpha_G
=Gs$, so that for sufficiently large values of the energy, namely, larger than the
Planck energy, the gravitational quantum fluctuations become large.
We see that in FQFT gravity this may not be the
case, because the finite quantum loop corrections become negligible in the high energy limit.
Of course, the contributions of the tree graph exchanges of virtual gravitons can be 
large in the high energy limit, corresponding to strong classical gravitational fields.
It follows that for high enough energies, the quantum behavior of fields propagating in a classical
curved spacetime would be a good approximation.

\section{Black Hole Entropy}

By using our perturbatively finite quantum gravity theory, we can investigate the 
Bekenstein-Hawking (BH) black hole entropy formula\cite{Bekenstein,Hawking}:
\begin{equation}
S_{\rm BH}=\frac{A_H}{4\hbar G},
\end{equation}
where $A_H$ is the black hole horizon. We shall adopt a conical Rindler space approximation to
a Schwarzschild black hole and obtain from the classical tree graph level Einstein-Hilbert action
the partition function\protect\cite{Susskind,Iellici}:
\begin{equation}
S_{\rm BH}=-\beta^{-2}\partial_\beta[\beta^{-1}\ln(Z)]=\frac{4\pi
GM^2}{\hbar}=\frac{A_H}{4\hbar G},
\end{equation}
where $\beta=T^{-1}$ is the inverse temperature and $M$ is the mass of the Schwarzschild 
black hole. Quantum corrections come from scalar
field fluctuations at the horizon and scalar field-graviton loop corrections.

In FQFT the scalar field first order self-energy loop contribution to the entropy is given by
\begin{equation}
S_{\phi}=\frac{A_H}{24\beta}m^2\Gamma\biggl(-1, \frac{m^2}{\Lambda_G^2}\biggr).
\end{equation}
For $m\rightarrow 0$, we get the finite result
\begin{equation}
S_{\phi}=\frac{A_H\Lambda^2_G}{24\beta}.
\end{equation}
The renormalized gravitational coupling constant $G_R$ redefines the entropy $S_{\rm BH}$
\protect\cite{Susskind,Iellici}:
\begin{equation}
S_{\rm BH}=\frac{A_H}{4\hbar G_R},
\end{equation}
where
\begin{equation}
\frac{1}{G_R}=\frac{1}{G_0}\biggl(1+\frac{\hbar G_0\Lambda_G^2}{6\beta}\biggr)
\end{equation}
and $G_0$ is the bare gravitational coupling constant.

For matter-graviton scattering, we obtain from FQFT gravity the
following contributions to the renormalized gravitational coupling constant:
\[
{\rm Loop\,Graphs}+{\rm Shadow\,Loop\,Graphs}+{\rm Ghost\,Loops}
\]
\[
+{\rm Shadow\,Ghost\,Loops}+{\rm Measure\,Loops}.
\]
This yields the renormalized $G$:
\begin{equation}
\frac{1}{G_R}=\frac{1}{G_0}[1-I_1(p^2=\mu^2)+I_2(p^2=\mu^2)+...],
\end{equation}
where $I_i(p^2=\mu^2)$ are the finite loop corrections evaluated at $p^2=\mu^2$ at every order. 

From these results, we see that the entropy, $S_{\rm BH}$, is finite at the black hole event
horizon. This would not be true in the standard unrenormalizable quantum gravity theory, since
$G_R$ is not physically meaningful in D=4 spacetime. 

\section{Conclusions}

We have formulated a perturbatively finite, unitary and gauge invariant field theory.
All the tree graphs for the gauge theory are local and 
yield the same predictions as the tree graphs in standard local field theory. If we take the
regulating limit $\Lambda\rightarrow\infty$, then we obtain the standard local field theory
together with an unrenormalizable gravity theory. On the other hand, if we choose $\Lambda_G$
to have a finite, fixed value, e.g., the Planck mass, then the resulting 
quantum gravity theory is perturbatively finite, gauge invariant under generalized local 
$SO(3,1)$ gauge tranformations, diffeomorphism invariant and unitary to all orders.
Thus, the FQFT formalism provides a consistent quantum gravity theory at least in the
perturbative regime. Since the graviton tree graphs are purely local they reproduce
the classical GR theory in the low-energy region together with the standard experimental
agreement of GR. The extended nonlinear gauge symmetry guarantees the decoupling of all
unphysical modes in the theory, using the shadow field formalism and the functional integral
technique together with the BRS ghost formalism, which can be consistently incorporated in the
shadow field formalism.

Within the quantum gravity theory, we can obtain a physically meaningful renormalization
of the gravitational coupling constant, which gives a finite value for the Bekenstein-Hawking
entropy formula at the event horizon. This could lead to a possible resolution of Hawking's
information loss problem\protect\cite{Hawking2}, although the issue of how the finite quantum
fluctuations modify the Hawking radiation Planckian spectrum remains to be fully resolved.
\vskip 0.2 true in
{\bf Acknowledgments}
\vskip 0.2 true in
This work was supported by the Natural Sciences and Engineering Research Council of Canada.
\vskip 0.5 true in

\end{document}